# Gate Voltage Control of Transition Metal Dichalcogenide Monolayers Quantum Yield

**Maksym V. Strikha[1,2], Anatolii I. Kurchak[2], and Anna N. Morozovska[3*]**

[1] *Taras Shevchenko Kyiv National University, Faculty of RECS,*

[2]*V.Lashkariov Institute of Semiconductor Physics, NAS of Ukraine,*

[3]*Institute of Physics, NAS of Ukraine.*

Two-dimensional transition metal dichalcogenide (2D-TMD) monolayers, which reveal remarkable semiconductor properties, are the subject of active experimental research. It should be noted that, unlike bulk TDMs, which are indirect-band semiconductors, 2D-TMD monolayers have the extreme points of the conduction and valence bands at the same K and K' points of the Brillouin zone. Therefore they are direct-band semiconductors and can claim to be widely used in optoelectronics devices. Recently it has been shown experimentally that quantum yield in $MoS_2$ and $WSe_2$ monoatomic layers can reach values close to unity when electrostatic doping makes them intrinsic semiconductors. However, the available theoretical description does not give an understanding of the physical mechanisms underlying in the gate voltage control of quantum yield.

This work is an attempt to propose a consistent semi-phenomenological theory of photo-induced charge carriers relaxation in 2D-TMDs, which allows obtaining an analytical dependence of the quantum yield on the voltage applied to the FET gate. We consider a standard experimental situation, when the 2D-TMD monolayer and the metal gate are plates of a flat capacitor, and the capacitor charge is proportional to the gate voltage. The dependences of the TMD monolayers quantum yield on the gate voltage and the carrier generation rate have been calculated for the cases of the prevailing recombination of free electrons and holes (radiative and non-radiative Auger recombination) and recombination of excitons (radiative and Auger recombination). In both cases analytical expressions were derived for the dependence of quantum yield on the gate voltage and photo-induced carriers generation rate at a fixed gate voltage. Quantitative agreement with experiment allows concluding about the relevance of the proposed theoretical model for the description of carriers photo-generation and recombination in 2D-TMD monolayers. Obtained results demonstrate the possibilities of 2D-TMD quantum yield control by the gate voltage and indicate that 2D-TMDs are promising candidates for modern optoelectronics devices.

---

[*] Corresponding author, e-mail: anna.n.morozovska@gmail.com

# I. INTRODUCTION

In recent years, there has been an active study of two-dimensional transition metal dichalcogenides (2D-TMDs), such as $MoS_2$, $WSe_2$, $MoTe_2$ [1, 2, 3]. A lot of theoretical and experimental studies are devoted to the band structure and electronic properties of 2D-TMDs [4, 5, 6, 7, 8, 9, 10], which demonstrate a possibility to tune the properties in a wide range due to mechanical strain [4, 5], or/and additional impurities [10]. In particular, the first direct observation of the transition from indirect to direct band-gap state has been reported in monolayer $MoSe_2$ [6]. Possible applications of 2D-TMDs was studied in combination with bilayer graphene [11] (possibility to tune electronic structure of $MoS_2$) and as 2D hybrid semiconductor-ferroelectric structure [12] (ultra fast, non-volatile multilevel memory devices with non-destructive low-power readout).

Two-dimensional transition metal dichalcogenides, such as $MoS_2$, $WSe_2$, unlike semi-metallic graphene, have semiconductor properties, and unlike indirect gap bulk TDM, are direct gap semiconductors, and therefore can be widely used in optoelectronics devices (see, e.g. Refs. [13, 14, 15]). The theory of photo-carriers relaxation in 2D-TMDs was developed by Kozawa et al. [16]. They assumed that the quantum yield (**QY**) in 2D-TMDs is low due to the separation by internal field of the photo-induced electrons and holes to different points in the Brillouin zone (electrons come to the $\Lambda$-point and holes come to the-$\Gamma$-point). The separation reduces the probability of the carriers' radiative recombination in the K-point, where the top of the valence band and the bottom of the conduction band are located. However, when the carriers are excited by light quants of relatively low energy, only a small fraction of them are photo-induced in the part of the Brillouin zone, where a "band-nesting" effect is possible and hence the QY values may be higher.

Recently it has been experimentally shown [17] that the QY in monoatomic layers of $MoS_2$ and $WSe_2$ can reach values close to unity under electrostatic doping of these materials to the intrinsic semiconductor state. However, the theoretical description performed in [17] being relevant per se, at the same time does not give clear understanding why the QY is strongly dependent on the gate voltage and what happens with its changes. Therefore, the aim of our work is to construct a consistent semi-phenomenological theory of the photo-induced carriers relaxation in 2D-TMDs, which allow to obtain the analytical dependence of the QY value on the voltage applied to the gate of field effect transistor (**FET**).

# II. PROBLEM FORMULATION

Here we consider a standard experimental situation. A 2D-TMD monolayer and a metal gate are the plates of a flat capacitor, and the charge of each plate per unit area is equal to

$$Q = C_{ox}(V_g - V_T), \quad (1)$$

where $C_{ox}$ is the specific capacitance per unit area determined by the dielectric constant of the substrate and its thickness; $V_g$ is the gate voltage. The presence of $V_g$ in Eq. (1) reflects the fact that doping defects are always present in 2D-TMDs. For instance 2D-TMDs studied in Ref.[17] had n-type conductivity, and it was necessary to apply a negative gate voltage –20 V to transform them to the intrinsic semiconductor state. Considering further that the pristine 2D-TMD is electronic (that corresponds to the standard experiments [15]), we write a simple expression for "doping" voltage $V_T$:

$$V_T = \frac{eN_D}{C_{ox}}, \quad (2)$$

here $e$ is the elementary charge, $N_D$ is the concentration of ionized donors. The geometry of the situation under consideration is shown in **Fig. 1**.

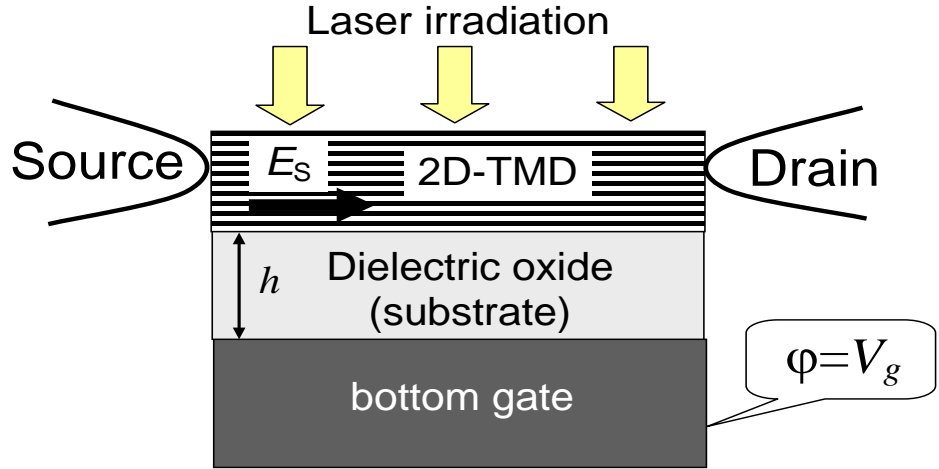

**Fig. 1.** Structure of the FET consisting of 2D-TMD monolayer - dielectric oxide - bottom gate. The 2D-TMD channel is stretched between the source and drain electrodes. Arrows indicate that a laser radiation exposes the surface of the 2D-TMD monolayer.

The 2D MoS$_2$ and WSe$_2$ monolayers, in contrast to corresponding indirect gap bulk materials not suitable for optoelectronics, are direct-gap semiconductors with an optical gap $E_g$ ~ 1.8 eV, respectively, and with extrema of the conduction and valence bands in points $K, K'$ of the hexagonal Brillouin zone [14, 15], just as in graphene. The spectrum of the conduction band of these materials also contains a side extremum located in the direction from the points $K, K'$ to the center Γ of the Brillouin zone. The presence of two sub-bands in the conduction band causes the possibility to realize the negative differential conductivity in the 2D WSe$_2$ and MoS$_2$. The effect is associated with the occupation of a higher valley with a higher effective mass by hot electrons (see theory proposed in Ref.[18]).

Two important features of the 2D-TMD band structure should be emphasized. The spin degeneration of the valence band at the points *K, K'* is removed by the spin-orbital interaction, but remains for the bottom of the conduction band [14, 15]. The magnitude of the spin splitting of the valence band top in the 2D-TMD is quite significant (about 160 meV for MoS$_2$ and 180 meV for MoSe$_2$), therefore we can assume that only the upper valence subband is occupied by holes at room temperature. The second feature is that the Coulomb interaction between electrons and holes increases significantly in the monoatomic TMD layers [14], and the optical gap becomes significantly smaller than the electronic one, which appears in the expressions for electron and hole concentrations. For MoS$_2$, the exciton binding energy is at least 0.48 eV [14], and therefore the electronic gap is about 2.3 eV.

In what following we consider the relatively low levels of photo-generation, when the equilibrium and non-equilibrium carrier concentrations are:

$$n = n_o + \delta n, \quad p = p_o + \delta p, \tag{3a}$$

$$n_o \gg |\delta n| \quad p_o \gg |\delta p|. \tag{3b}$$

We will assume that the equilibrium carriers are non-degenerated. The assumption corresponds to the experimental situation [17], when the Fermi level is located deep enough in the band gap. Taking into account the double valley degeneration in 2D-TMDs (see, e.g., review [19] and refs. therein) the concentrations of non-degenerate equilibrium 2D electrons and holes are:

$$n_o = \frac{2m_c}{\pi\hbar^2} e^{-\frac{E_c - E_F}{kT}} \equiv N_c e^{-\frac{E_c - E_F}{kT}}, \quad p_o = \frac{m_v}{\pi\hbar^2} e^{-\frac{E_F - E_v}{kT}} \equiv N_v e^{-\frac{E_F - E_v}{kT}} \tag{4}$$

Here $E_{c,v}$ are the energies of the conduction and the valence band edges at *K* and *K'* points, $E_F$ is the Fermi level energy, $N_{c,v}$ are the effective densities of states in the conduction and valence bands, $m_{c,v} = \sqrt{m_{c,v}^{\parallel} m_{c,v}^{\perp}}$ are the averaged effective masses of carriers in the anisotropic conduction and valence bands, respectively. Note that the values $N_c$ and $N_v$ in Eq.(4) also differ by the factor 2, which is caused by the above mentioned removal of the spin degeneracy of the valence band top. A standard expression inherent to bulk semiconductors follows from Eq.(4):

$$n_o p_o = N_c N_v e^{-\frac{E_g}{kT}} \equiv n_i^2. \tag{5}$$

Here $n_i$ is the carrier concentration in the intrinsic semiconductor, which Fermi level is located in the middle of the gap. It should be emphasized that Eq,(5) contains the electronic gap value, which is equal to the sum of the optical value and the binding energy of the exciton, $E_g = E_g^{opt} + E_{ex}$.

Next, we rewrite expression (1) as:

$$n - p = C_{ox}\frac{V_g - V_T}{e}. \qquad (6)$$

Using expressions (5)-(6) and taking into account the inequalities (3b), we obtain the quadratic equation, $np = n\left(n - C_{ox}\dfrac{V_g - V_T}{e}\right) = n_i^2$, which solution is:

$$n = \frac{C_{ox}(V_g - V_T)}{2e} + \sqrt{\left(\frac{C_{ox}(V_g - V_T)}{2e}\right)^2 + n_i^2}, \qquad (7a)$$

$$p = \frac{n_i^2}{n} \equiv -\frac{C_{ox}(V_g - V_T)}{2e} + \sqrt{\left(\frac{C_{ox}(V_g - V_T)}{2e}\right)^2 + n_i^2}. \qquad (7b)$$

From expressions (7), $n = p \approx n_i$ for $V_g = V_T$, and $n \approx C_{ox}V_g/e \gg p$ for $V_g \gg V_T$ corresponding to a sharply unipolar conductivity. Solutions for *p*-type TMD can be obtained from Eqs.(7), where the sign before the first term should be changed.

## III. QUANTUM YIELD FOR THE DOMINANT RECOMBINATION OF FREE ELECTRONS AND HOLES

In what following, we consider separately two cases of recombination; the first when almost all photo-induced carriers are free, and the second, when almost all photo-induced carriers at first bind into electrically neutral excitons and subsequently recombine. It is clear that the first case takes place at sufficiently high temperatures, $kT \gg E_{ex}$, and the second case corresponds to relatively low temperatures, $kT \ll E_{ex}$. Generally speaking, in the second case, one should consider the binding of photo-induced carriers into trions and bi-excitons as well, but since their binding energy is much smaller than the binding energy of excitons we neglect these possibilities. In particular, for MoS$_2$, the binding energy of the trion is about 18 meV [14], that is slightly lower than the thermal energy at room temperature. Hence, without loss of generality, we will assume that the concentration of trions and bi-excitons is much smaller than the concentration of excitons, and so they will not significantly affect other channels of recombination. It should also be noted that, due to the large exciton binding energy values in the 2D-TMD monolayers, the second case takes place for the vast majority of practical situations. However, for the sake of generality, we consider both cases in the sequence indicated above.

In what following, we consider two major relaxation channels of free photo-carriers, the radiative recombination, which rate is proportional to the product of the electrons and holes concentrations (the recombination energy is taken away by the light quantum), and the non-radiative Auger recombination, when the recombination energy transfers to the third carrier through Coulomb interaction, and the carrier goes into an excited state in the band structure and then relaxes, transferring energy to the lattice phonons or substrate (**Fig. 2**). Also we neglect other recombination mechanisms, such as Shockley-Reed recombination through defects, because of their lower intensity [16]. For low levels of quantum generation, which energy does not much exceed the gap width, the effect of the photo-induced electrons and holes "nestling" can also be neglected [16].

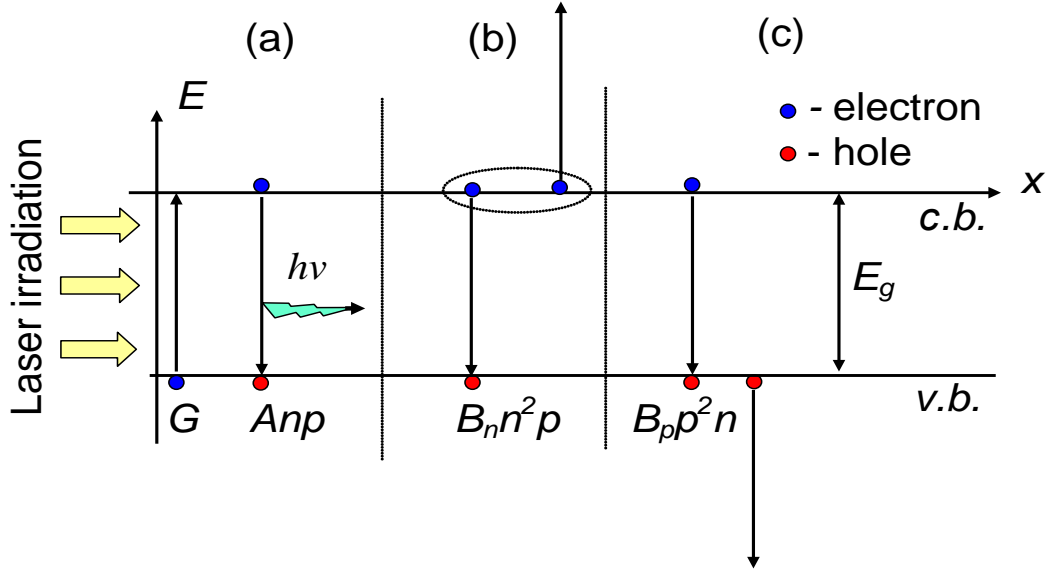

**Fig. 2.** Recombination processes in 2D-TMDs. **(a)** Radiative exciton recombination; **(b-c)** non-radiative Auger recombination of an electron-hole pair with energy transfer to a third electron **(b)** or hole **(c)**. A Coulomb interaction between the electrons inside the oval transmits the recombination energy.

By definition, the QY equals to the ratio of the radiative recombination rate (number of recombination acts per unit time in a unit volume) to the total rate of radiative and non-radiative recombinations:

$$QY = \frac{T_R}{T_R + T_{NR}} \approx \frac{Anp}{Anp + B_n n^2 p + B_p np^2} \qquad (8)$$

Here, $A$ is the coefficient of radiative recombination, $B_{n,p}$ are the coefficients of the non-radiative Auger recombination involving the second electron or hole, respectively. It may seem that non-radiative Auger recombination is a priori less efficient in sufficiently wide gap semiconductors, such as 2D-TMDs. The fact is that the simultaneous implementation of the energy and momentum conservation laws imposes the process threshold; and so it is possible only when the total kinetic energy of the three carriers involved in it is higher than a certain threshold energy, $E_T$, and therefore the Auger recombination coefficient $B$ contains an exponential dependence $B \sim e^{-\frac{E_T}{kT}}$ [20]. Assuming parabolic bands with effective mass ratios $m_v \gg m_c$ it is easy to show that $E_T = \frac{m_c}{m_v} E_g$. Therefore, the threshold energy can be orders of magnitude higher than 0.026 eV, in a wide gap semiconductor, and so the intensity of Auger-recombination is extremely low even at room temperature. But the threshold $E_T$ can be significantly reduced due to the peculiarities of the band structure. Namely, if another parabolic valley with an effective mass $m_c^\Delta$ exists above the bottom $E_c$ of the conduction band, and the energy gap $\Delta E \leq E_g$ separates its from the bottom (see **Fig. 3b**), it is easy to show that the threshold energy is $E_T = \frac{m_c^\Delta}{m_v}(E_g - \Delta E)$, and it can be small enough if $\Delta E$ is close to $E_g$. The calculations [4, 13-15] show that the 2D-TMD band structure contains upper valleys for both the conduction and valence bands meeting the requirement, $E_g - \Delta E \ll E_g$, and therefore we can expect high coefficients $B_{n,p}$ in these materials. This means that Auger-recombination can be intensive enough and significantly reduce the QY for high concentrations of carriers.

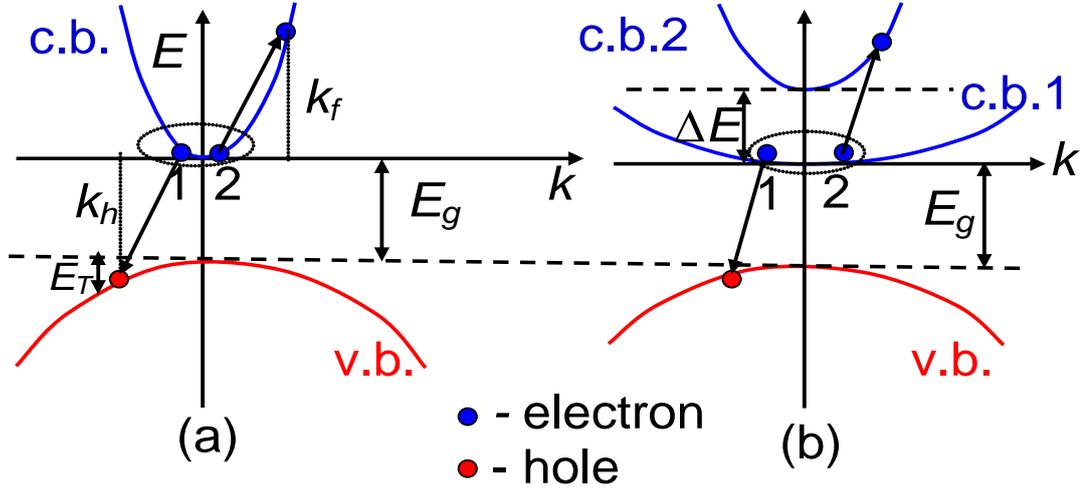

**Fig. 3.** Threshold for Auger recombination in the system with two **(a)** and three **(b)** bands. Coulomb interaction transmits the recombination energy of the order of the gap width between the electrons 1 and 2.

The expressions $E_T \approx \frac{\hbar^2 k_h^2}{2m_v}$ and $|k_h| \approx \frac{\sqrt{2m_c E_g}}{\hbar}$ are true for figure (a), and the expressions $|k_h| \approx |k_f|$ and $|k_h| \approx \frac{\sqrt{2m_c^\Delta (E_g - \Delta E)}}{\hbar}$ are valid for figure (b) from the energy and momentum conservation laws.

We can rewrite expression (8) for QY as a function of electron concentration $n$:

$$QY[n] = \frac{1}{1 + \frac{B_n}{A}(n + \beta p)} \approx \frac{1}{1 + \alpha\left(n + \beta \frac{n_i^2}{n}\right)}, \qquad (9)$$

where the parameters $\alpha = \dfrac{B_n}{A}$ and $\beta = \dfrac{B_p}{B_n}$ are introduced. As can be seen from expression (9), the value of the QY reaches a maximum for $\beta=1$, exactly when the semiconductor becomes the intrinsic one under corresponding gate voltage . Knowing the electron and/or hole lifetimes for different recombination processes it is possible to estimate the parameters $\alpha$ and $\beta$. The values of $N_d$ and $n_i$ are determined by the 2D-TMD sintering technology and ambience temperature; and it can vary over a wide range of values.

Note that according to Eqs.(6)-(7), the carrier concentrations $n$, $p$ and their difference $(n-p)$ depend on the voltage difference $V_g - V_T$ only, so it makes sense to plot the concentration and QY as functions of $V_g - V_T$.

**Figures 4** show for the dependence of $n$, $p$ and QY of the 2D-TMD on the gate voltage, calculated numerically using Eqs. (7) and (9) for parameters $\beta=1$, $\alpha=10^{-12}$ m$^2$ and $n_i = 10^8$ m$^{-2}$, which order of magnitude corresponds to the MoS$_2$ electronic gap at room temperature, and for the capacitance values $C_{ox}=(200 – 2)$ nF/m$^2$, which reflect the wide variation of the dielectric constant and the thickness of underlying substrate.

**Figure 4a** shows that the concentration of electrons is equal to the concentration of holes at $V_g = V_T$, and, the concentration of carriers (electrons or holes depending on the sign $V_g - V_T$) begins to increase sharply when $|V_g - V_T|$ increases. Holes dominate for $V_g - V_T < 0$ and electrons dominate for $V_g - V_T > 0$. The order of magnitude of the carrier concentration increases sharply with the increase of $C_{ox}$, reflecting the fact that the high substrate capacity greatly contribute to the selectivity of the FET, and, consequently, increases the QY of the 2D-TMD.

Indeed, it is seen from **Fig. 4b** that the QY being close to 100% at $V_g = V_T$, decreases symmetrically with increasing $|V_g - V_T|$, since concentration of carriers increases sharply with increasing $|V_g - V_T|$. The sharpness of the QY maxima at $V_g = V_T$ increases significantly with increasing $C_{ox}$.

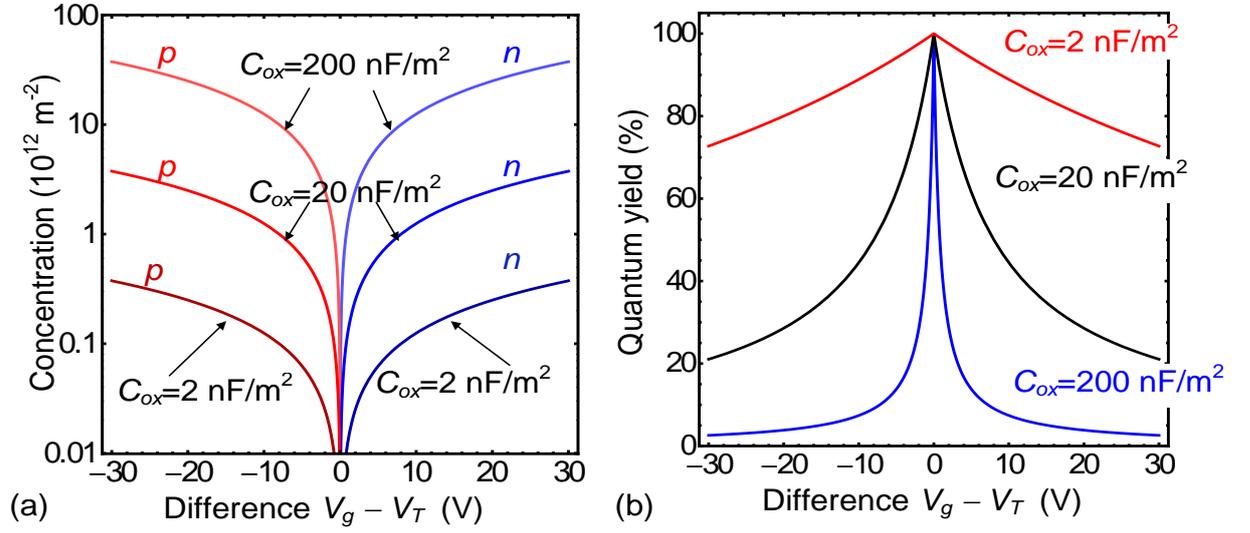

**Fig. 4.** The dependence of the 2D-TMD carrier concentration **(a)** and QY **(b)** on the voltage difference $V_g - V_T$, calculated using Eqs.(7) and (9), respectively. Capacitance $C_{ox}$ = 200, 20 and 2 nF / m² for different curves, $\alpha = 10^{-12}$, $n_i = 10^8$ m⁻² and $\beta = 1$.

**Figure 5** shows the dependence of the TMD QY on the gate voltage $V_g$ and parameter $\beta$ calculated from Eq.(9), for the same values of parameters $\alpha$ and $n_i$ as in **Fig. 4**. Comparing **Fig. 5a** and **5b**, calculated for relatively small (2 nF/m²) and bigger (20 nF /m²) $C_{ox}$, we see that the selectivity of the 2D-TMD is significantly improved with increasing $C_{ox}$. Note that the maximum QY corresponds to a difference of $|V_g - V_T| < 1$ V. QY decreases with increasing $|V_g - V_T|$, but the increase depends significantly on $\beta$ and is asymmetric with respect to $|V_g - V_T|$. In particular, when $\beta \ll 1$ a much higher QY corresponds to negative values $V_g - V_T$, and the opposite situation takes place at $\beta \gg 1$, due to the different intensities of Auger recombination involving electron and hole. The sharpness of the dependences on $QY(V_g - V_T)$ increases with increase of $C_{ox}$, as can be seen from the comparison of **Figs. 5a** and **5b**.

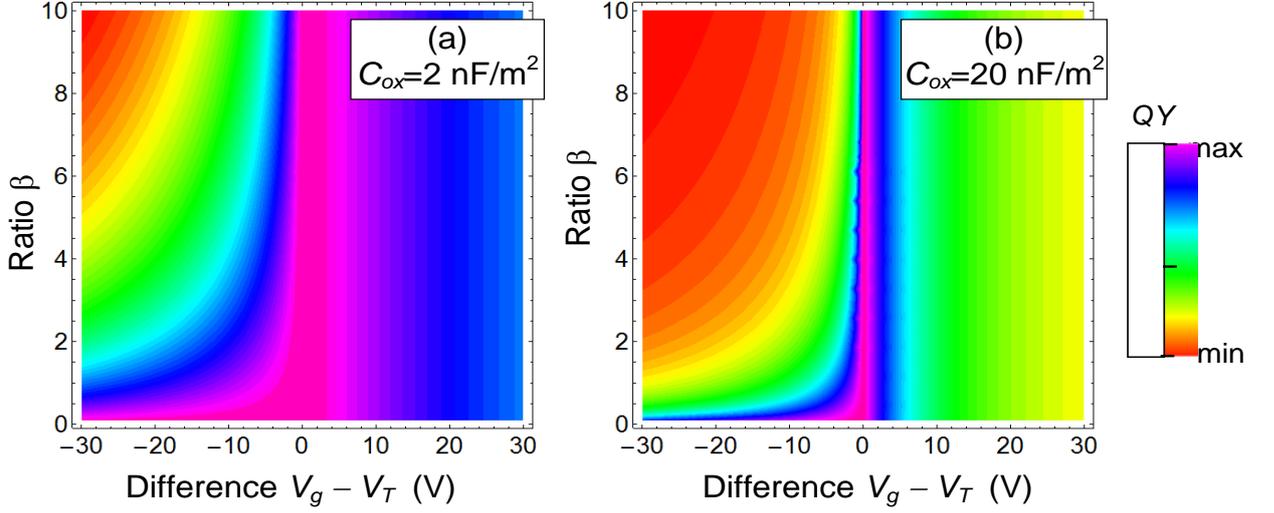

**Fig. 5.** Dependence of the 2D-TMD QY on the voltage difference ($V_g - V_T$), calculated using Eq.(9), for capacitance $C_{ox}$=2 nF/m² (a) and 20 nF/m² (b). Parameter values are $\alpha=10^{-12}$ m² and $n_i = 10^8$ m⁻².

To obtain the dependence of the QY on the photo-induced carriers generation rate $G$, we use the following equation of the carrier balance:

$$\frac{\partial n}{\partial t} = G + (\Delta - Anp) + (\Gamma_n n - B_n n^2 p) + (\Gamma_p p - B_p np^2). \qquad (10)$$

Here, in brackets, the rates of the recombination processes and corresponding inverse ionization processes are collected in pairs [20]. It should be noted that one of the important shortcomings of the theoretical consideration [17] is the complete ignorance of inverse processes. Under steady-state and thermal equilibrium conditions, $G = 0$ and $\frac{\partial n}{\partial t} = 0$. So that each bracket in the right-hand side of Eq.(10) must be zero, namely:

$$\Delta = An_i^2, \quad \Gamma_n = B_n n_i^2, \quad \Gamma_p = B_p n_i^2. \qquad (11)$$

Therefore, for stationary conditions we obtain the simple expression for the QY as a function of carrier concentrations and generation rate,

$$QY[n, p, G] \approx \frac{Anp}{G + (A + B_n n + B_p p)n_i^2} \approx \frac{1}{\frac{G}{An_i^2} + 1 + \alpha\left(n + \beta\frac{n_i^2}{n}\right)}. \qquad (12)$$

The approximate equality in Eq.(12) is valid at $np \approx n_i^2$.

For the "conditional" comparison with experiment [17], expression (12) should take into account that the rate $G$ is no longer dependent on $n$ and $p$, since the recombination model (12) is valid at sufficiently high temperatures for which corresponding thermal energy exceeds significantly the exciton binding energy, $kT >> E_{ex}$, and so only free carriers can recombine. The calculation results according to Eq.(12) is shown in **Appendix, Fig. A1**. The solid curves in this figure quantitatively describe the experimental results [17], but they are plotted for parameter values $\alpha=6.7\times10^8$ m$^2$ and $\beta=50$, which are obviously not realistic. This should not raise questions because of the conditionality of the comparison.

## IV. QUANTUM YIELD UNDER EXCITON RECOMBINATION

As we saw above, in fact, for all real monoatomic 2D-TMD layers, the exciton binding energy is very high and the inequality $kT << E_{ex}$ is valid. Therefore, almost all free carriers, which can find a pair, bound into excitons. The concentration of excitons $n_{ex}$ will be approximately equal to the concentrations of minority carriers, i.e. electrons in *p*-type material and holes in *n*-type material.

In this case, the physical picture of recombination processes depicted in **Fig. 2** undergoes changes. In each case, not the free electrons and holes recombine, but only the electrons and holes bound in the excitons recombine, and therefore the recombination rate is no longer proportional to the product *np*, but to the concentration $n_{ex}$. The rate of non-radiative Auger recombination appeared proportional also to the concentration of majority carriers (electrons or holes), which are not bound in excitons, and obtain the energy of the gap width order through Coulomb interaction. Since the exciton wave function contains a complete set of momenta, the Auger-recombination process involving an exciton has no energy threshold, which increases its probability [20].

The correct expression for QY takes into account that almost all minority carriers are bound into excitons in the low-temperature approximation, and there are virtually no minority carriers capable to take away recombination energy; there are only majority carriers that can do this. Therefore, the Auger recombination process with the participation of a free electron only can occur in the *n*-type 2D-TMD, and the process with the participation of a free hole only can exist in the *p*-type 2D-TMD. This leads to the expressions:

$$QY_{(n)} = \frac{T_R}{T_R + T_{NR}^{(n)}} \approx \frac{A^{ex} n_{ex}}{A^{ex} n_{ex} + B_n^{ex} n_{ex}(n - n_{ex})}, \qquad (13a)$$

$$QY_{(p)} = \frac{T_R}{T_R + T_{NR}^{(p)}} \approx \frac{A^{ex} n_{ex}}{A^{ex} n_{ex} + B_p^{ex} n_{ex}(p - n_{ex})}. \qquad (13b)$$

Here $A^{ex}$ is the coefficient of radiative exciton recombination, $B_{n,p}^{ex}$ are the coefficients of the nonradiative Auger recombination of exciton with the participation of the second electron or hole, respectively. Note that the dimensions of the coefficients $A^{ex}$ and $B_{n,p}^{ex}$ in Eq.(12) are different from the dimensions of the coefficients $A$ and $B_{n,p}$ in Eq.(12). Taking into consideration that for *n*-type 2D-TMD the concentration of excitons is approximately equal to the concentration of minority holes, $n_{ex} \approx p$, and for p-type 2D-TMD $n_{ex}$ is defined by the concentration of minority electrons, $n_{ex} \approx n$, we rewrite expressions (13) as:

$$QY_{(n)} \approx \frac{1}{1+(n-p)/N_n} \quad (14a)$$

$$QY_{(p)} \approx \frac{1}{1+(p-n)/N_p} \quad (14b)$$

The "characteristic" concentrations $N_n = \frac{A^{ex}}{B_n^{ex}}$ and $N_p = \frac{A^{ex}}{B_p^{ex}}$ are introduced. Note that the values $N_{n,p}$ have the dimension of 2D concentration in SI units.

An important consequence of expressions (14) is that the QY does not depend on the concentration of excitons, but is sensitive to the concentration of free carriers only. Expressions (14) also show that the QY reaches a maximum (100%) at $n = p$ corresponding to the intrinsic semiconductor state. The QY values smaller than 100% observed in the experiment [17] even for the case of the intrinsic semiconductor, can be explained by the fact that there are always other non-radiative recombination channels (e.g., through impurity centers), which are not taken into our simplified consideration.

**Figures 6-7** show the voltage dependences of QY calculated from Eqs.(14). The interpolation function $QY_{(n)} \approx QY_{(n)}UnitStep[n-p] + QY_{(p)}UnitStep[p-n]$, where $UnitStep[x]$ is the Heaviside step-function, was used to generate the plots. The interpolation is "tailoring" of Eqs.(14a) and (14b), that is continuous at $n = p$ for arbitrary concentrations $N_n$ and $N_p$. To simplify numerical analysis we assume that $N_n = N_p = N_{ex}$.

Note that according to Eqs.(6)-(7) the carrier concentrations and their difference $(n-p)$ depend only on the voltage difference $V_g - V_T$, so it makes sense to plot the *n, p* and QY in dependence on $V_g - V_T$. However, to demonstrate the separation of QY maxima, we show the dependencies $QY[V_g]$ calculated for different doping voltages $V_T$ in **Fig. 6b-d**.

**Figure 6a** shows the dependences of the *n* and *p* on $V_g - V_T$ calculated from Eq.(7) for the capacitance $C_{ox} = 20$ nF/m$^2$. The plot differs from the middle curve in **Fig. 4a** only by a linear scale of the vertical axis, and is shown here for the better understanding of QY behavior presented in **Figs. 6b-d.**

**Figures 6b-d** show the QY dependences calculated from Eqs.(14) on the gate voltage $V_g$ for different values of the doping voltage $V_T$ and effective concentration $N_{ex}$. As one can see from **Fig. 6b**, for a relatively small $N_{ex}$, the QY reaches values close to 100% only in a very narrow

vicinity of $V_g = V_T$. When $N_{ex}$ increases, the QY becomes higher in a wider range of gate voltages (see **Fig. 6c-d**). Physically, this is explained by the fact that the Auger recombination rate is substantially lower than the radiative recombination rate in the limiting case of high $N_{ex}$, that significantly increases QY.

The dependence of 2D-TMD QY on the voltage difference ($V_g - V_T$) and effective concentration $N_{ex}$ calculated using Eqs.(14) for $C_{ox}= 2$ nF/m$^2$ and $C_{ox}= 20$ nF/m$^2$ is shown in **Figs. 7a** and **7b**, respectively. It can be seen that the range of high QY increases linearly with $N_{ex}$ increase. The voltage range of high QY is symmetric with respect to $V_g - V_T$; and QY decreases significantly with $C_{ox}$ increase. Results presented in **Fig. 7** show a sharper character of the symmetric QY maximum at $V_g = V_T$, compared to the asymmetric case shown in **Fig. 4.**

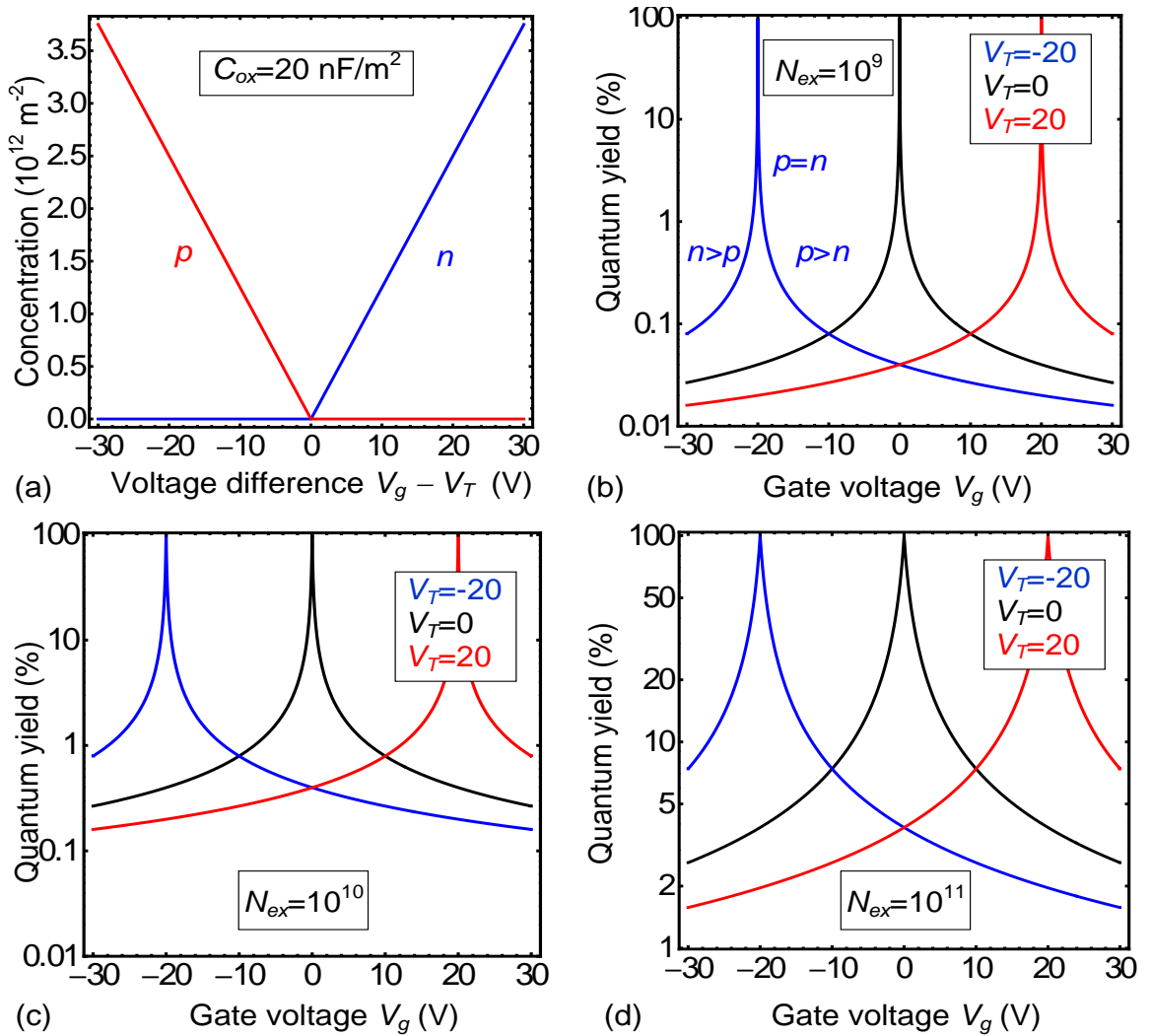

**Fig. 6. (a)** Dependence of carrier concentration calculated from Eq.(7) on voltage difference $V_g - V_T$ applied to 2D-TMD. **(b-d)** Dependence of the QY on the gate voltage calculated from (14) for several values of the doping voltage $V_T$= -20, 0, and 20 V (blue, black and red curves, respectively). Parameters $C_{ox}$ = 20 nF/m$^2$, $n_i$ = $10^8$m$^{-2}$ and $N_{ex}$ = $10^9$ m$^{-2}$ (b), $10^{10}$ m$^{-2}$ (c), and $10^{11}$ m$^{-2}$ (d).

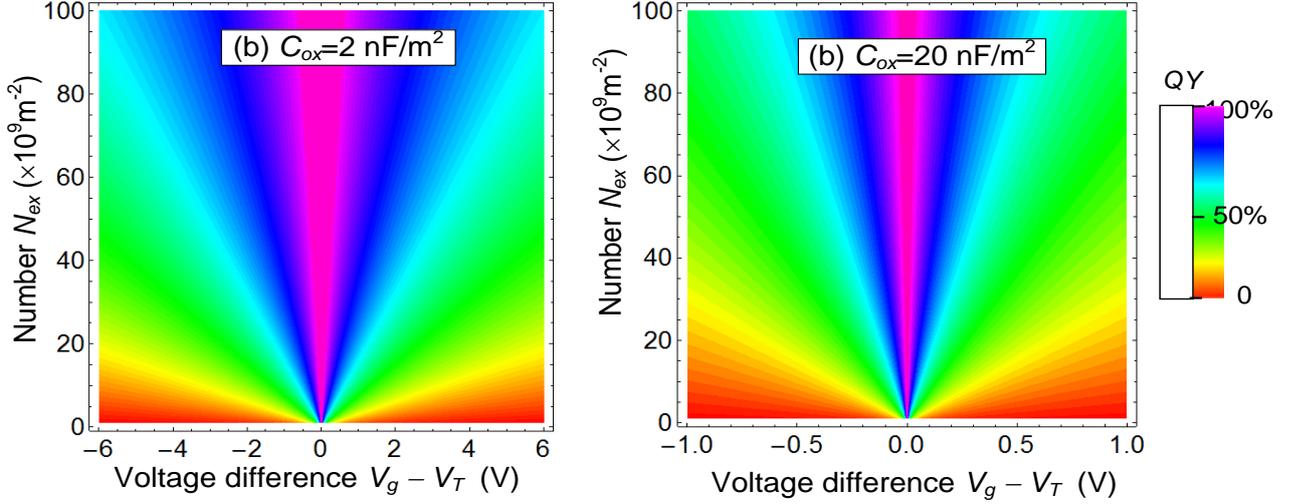

**Fig. 7.** Dependence of the 2D-TMD QY on the voltage difference ($V_g - V_T$) and the effective concentration $N_{ex}$ calculated from Eqs.(14) for the values $C_{ox}$ = 2 nF/m$^2$ **(a)** and $C_{ox}$ = 20 nF/m$^2$ **(b)**. Other parameters are the same as in **Fig. 6.**

For comparison with experiment [17], we calculate the dependence of the QY (14) on the generation rate $G$ at a fixed gate voltage, using approximate dependences of the concentrations of $n$ and $p$ on $G$.

To obtain the dependencies $n(G)$ and $p(G)$, we take into account that the difference of negative and positive charges is described by expression (6) only for the case of small generation rates $G$, when the relation (3) holds. Then the solutions (7) are valid, and the QY (14) does not depend on $G$.

For high generation rates Eq.(3a) is valid, and the inequalities (3b) do not hold, or even turns into inverse inequalities $n_o \leq |\delta n|$ and $p_o \leq |\delta p|$. If we consider the $n$-type 2D-TMD with $n_o > p_o$ and the generation rate is high enough, then $n_o \sim \delta n$. We regard that there are many trapping levels for minority carriers, and the carriers are effectively captured by traps, and so $\delta n \gg \delta p$ (see, e.g., [20]). That is, there are free photo-induced electrons, but actually no free photo-induced holes, because they are immediately captured at the traps, much faster that all other processes occur. As a consequence, the electron concentration $n$ in the denominator of Eq.(14a) is still described by Eq.(3a), but contains a non-equilibrium part that depends on $G$. The concentration of holes $p \approx p_o$, because the photo-induced carriers are immediately captured at the trapping levels, and cannot participate in the formation of excitons with free electrons. Of course, this is the simplest model of the recombination processes, because bound excitons and trions [17] can also be formed.

Thus, for the *n*-type doping the denominator in Eq.(14a) transforms as $n - p \to n_o + \delta n(G) - p_o$, and the dependence (14a) acquires the form

$$QY_{(n)}[G] \approx \frac{N_{ex}}{N_{ex} + \delta n(G) + C_{ox}\dfrac{V_g - V_T}{e}} \qquad (15a)$$

Equation (15a) is valid at $C_{ox}\dfrac{V_g - V_T}{e} \geq 0$. To compare it with experiment [17] it's necessary to know the dependence $\delta n(G)$. The simplest approximation is a linear function

$$\delta n(G) = \tau_{eff} G, \qquad (15b)$$

which includes an effective lifetime $\tau_{eff}$, that, in general, can be very different from all other recombination times [20]. If there are many recombination channels and trapping levels, a power approximation can be used instead of (15b),

$$\delta n(G) \cong N_T \left(\frac{\tau_{eff} G}{N_0}\right)^m, \qquad (15c)$$

where the factor $m$ can be different from unity and is slightly dependent on $V_g$. The dimensionality of parameters $N_T$ and $N_0$ is m$^{-2}$.

Similarly, *p*-type doping of 2D-TMD in the presence of trapping levels for minority electron carriers leads to the substitution $p - n \to p_o + \delta p(G) - n_o$ in the denominator (15b), and the dependence (14a) acquires the form

$$QY_{(p)}[G] \approx \frac{N_{ex}}{N_{ex} + \delta p(G) - C_{ox}\dfrac{V_g - V_T}{e}}. \qquad (15d)$$

Expression (15d) is valid at $C_{ox}\dfrac{V_g - V_T}{e} < 0$. Note that Eq.(15) are mathematically identical to formulas (12) at $m = 1$.

The dependence of QY on the generation rate $G$ at a fixed gate voltage $V_g$ is shown in **Fig. 8**. For the figure we used the interpolation function, $QY_{(n)} \approx QY_{(n)} UnitStep[n_o - p_o] + QY_{(p)} UnitStep[p_o - n_o]$, where *UnitStep*[$x$] is a Heaviside step-function, and consider two cases, $m = 1$ (**Fig. 8a**) and *m* is a fitting parameter (**Fig. 8b**).

The solid curves in both figures describe the experimental data [17] semi-quantitatively and correspond to realistic values of the fitting parameters, $V_T \approx 20$ V, $C_{ox}$=(15 - 50) nF/m$^2$, $N_{ex}$=(1 – 4)×10$^9$ m$^{-2}$, $m \approx 1$ or 0.7. To find the best agreement with experimental points [17], we varied the fitting parameters in a wide range, but reached the best fitting for the points at $V_g \approx 20$ V, since this case corresponds to the highest QY. The fitting of the experimental points at $V_g \approx 0$ and $V_g \approx -20$ V, corresponding to the much smaller QY looks worse. The fitting procedure involved a certain "weighting" of experimental points for each $V_g$, which were included into the list-squire functional and were adapted to minimize simultaneously the functional for three curves at $V_g = -20, 0$ and 20 V at the same fitting parameters.

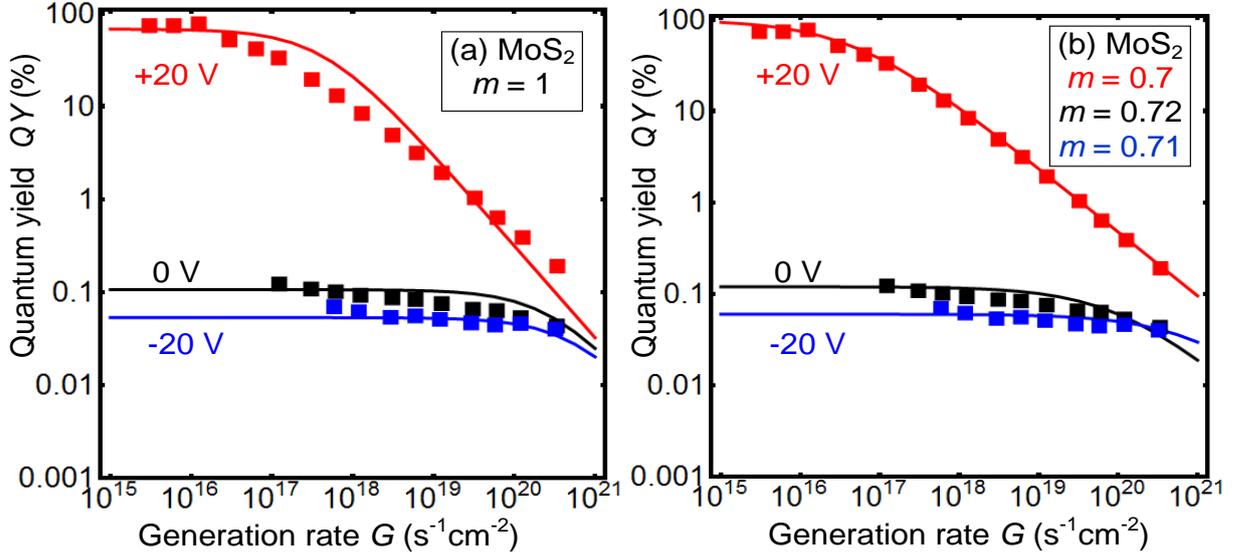

**Fig. 8.** The dependence of the QY on generation rate $G$ at several gate voltages. Symbols are experimental data [17] for the gate voltages +20 V (red), 0 (black) and -20 V (blue), red, black and blue solid curves are fitted by Eqs.(15) for parameters **(a)** $V_T \approx 20$ V, $C_{ox}=15$ nF/m$^2$, $N_{ex}=1\times10^9$ m$^{-2}$, $m\approx1$, $\tau_{eff}=5\times10^{-9}$s; **(b)** $V_T \approx 20$ V, $C_{ox}=50$ nF/m$^2$, $N_{ex}=3.75\times10^9$ m$^{-2}$, $m\approx0.7$, $N_T\left(\dfrac{\tau_{eff}}{N_0}\right)^m = 7.84\times10^{-3}$ SI units.

Note that the satisfactory accuracy of the fitting for all 3 curves obtained at $m=1$ and $m=0.7$ corroborates the relevance of the proposed theoretical model to the physical processes of carriers generation and recombination in 2D-TMD monolayers. A slight deviation of the parameter $m$ from the value 0.7 in **Fig. 8b** means that the trapping of carriers is rather weakly dependent on the gate voltage. The effect is quite possible. For example, the density of the screening charges in a paraelectric or ferroelectric substrate is nonlinearly dependent on the applied voltage, not only on the screened polarization [21, 22, 23]. Moreover, to achieve a sufficiently large capacitance $C_{ox}$ that increases the sensitivity of the FET (see **Figs. 6-7**), we may recommend the use of a ferroelectric or paraelectric perovskite oxides as a substrate.

Finally, let us consider the question of the QY dependence on generation rate in the context of more general model. The balance equation similar to Eq.(10) has the form,

$$\frac{\partial n}{\partial t} = G + (\Delta^{ex} - A^{ex}n_{ex}) + (\Gamma_n^{ex} - B_n^{ex}n_{ex})(n - n_{ex}) + (\Gamma_p^{ex} - B_p^{ex}n_{ex})(p - n_{ex}). \qquad (16)$$

Here, as in Eq.(10), the recombination and inverse ionization processes are gathered in brackets. Under the steady-state and thermal equilibrium conditions, each bracket in the right-hand side of Eq.(16) must also be zero. This gives:

$$\Delta^{ex} = A^{ex} n_{ex} \approx A^{ex}(p_o, n_o), \quad \Gamma^{ex}_{n,p} = B^{ex}_{n,p} n_{ex} \approx B^{ex}_{n,p}(p_o, n_o). \quad (17)$$

The concentration of intrinsic holes corresponds to the *n*-type 2D-TMD and the intrinsic electrons to the *p*-type TMD in the brackets of Eq.(16). Keeping this in mind the above considerations regarding the expressions (15), we can rewrite Eq.(16) for the *n*- and *p*-type TMD in the case of not too high photo-excitation:

$$\frac{\partial n}{\partial t} = G + A^{ex}(p_o - p) + \Gamma^{ex}_n (n - p_o)(p_o - p) = G - \left[A^{ex} + \Gamma^{ex}_n (n - p_o)\right]\delta p \equiv G - \frac{\delta p}{\tau^R_{ex}} - \frac{\delta p}{\tau^{NRn}_{ex}} \quad (18a)$$

$$\frac{\partial n}{\partial t} = G + A^{ex}(n_o - n) + \Gamma^{ex}_n (p - n_o)(n_o - n) = G - \left[A^{ex} + \Gamma^{ex}_p (p - n_o)\right]\delta n \equiv G - \frac{\delta n}{\tau^R_{ex}} - \frac{\delta n}{\tau^{NRp}_{ex}} \quad (18b)$$

Here we assume that Eq.(3) is still valid, and the characteristic times of radiative and non-radiative exciton annihilation are introduced as

$$\tau^R_{ex} = \frac{1}{A^{ex}}, \quad \tau^{NRn}_{ex} = \frac{1}{\Gamma^{ex}_n (n - p_o)}, \quad \tau^{NRp}_{ex} = \frac{1}{\Gamma^{ex}_p (p - n_o)}. \quad (19)$$

Allowing for Eqs.(16) and (19) in the stationary case, we can rewrite the expression for the QY for the *p*-type and *n*-type TMD as:

$$QY_{(n,p)} = \frac{\tau_{(n,p)}}{\tau^R_{ex}}, \quad \frac{1}{\tau_{(n,p)}} = \frac{1}{\tau^R_{ex}} + \frac{1}{\tau^{NR(n,p)}_{ex}}. \quad (20)$$

The total effective exciton annihilation times $\tau_{(n,p)}$ are introduced here. It is clear that when the exciton annihilation time is much smaller than the irradiation recombination time, the QY tends to unity. An important consequence of Eq.(20) is that the QY in the case of sufficiently low recombination rates does not depend on the generation rate *G*, but only on the concentrations of electrons and holes. However, as *G* increases and we go beyond the last of relations (3), because the concentrations in denominators (15) increase linearly with *G*, the QY decreases, which corresponds to the experimental dependences [17] (see **Fig. 8**).

## V. CONCLUSION

This work is the first attempt to propose the self-consistent semi-phenomenological theory of the photo-ionized carriers relaxation in 2D-TMDs, which allows deriving the analytical dependence of the QY on the voltage applied to the gate of the FET.

We considered the standard experimental situation when the 2D-TMD monolayer and the FET gate are plates of a flat capacitor, and the charge of each of plate is proportional to the gate voltage. The dependences of the QY on the gate voltage and intensity of carrier generation rate were calculated and analyzed.

The cases of dominant free electrons and holes recombination (radiative and non-radiative Auger recombination) and recombination of excitons (radiative and Auger recombination) are considered. We neglected other less intensive recombination channels, such as Shockley-Reed recombination, recombination of trions and bi-excitons, etc. For both cases (recombination of free carriers and recombination of excitons), analytical expressions have been derived for the concentrations of charge carriers and for the dependence of QY on the gate voltage, and the photo-carriers generation rate at fixed gate voltage.

It appeared that for real 2D-TMD monolayers the second case dominates, because the binding energy of excitons is very high (0.5 eV or more [14]) due to the strong electron-hole interaction in the monolayer, and therefore virtually all the minority carriers are bound into excitons at room temperature. Some of the majority carriers remain free, which allow Auger recombination by "taking away" the energy of the optical gap width order.

In both cases of free carriers and excitons recombination, the value of the QY reaches the maximum (order of unity) for the gate voltages corresponding to the intrinsic semiconductor state of 2D-TMD monolayer (i.e. when the concentration of free electrons is equal to the concentration of free holes). In the second case the reason for this is particularly physically transparent, because the free carriers capable to provide a non-radiative Auger recombination process are practically absent, and the recombination process occurs only through the radiative channel, except for Shockley-Reed recombination processes of lower intensity.

Quantitative agreement with experiment [17] (shown in **Fig. 8**) allows to conclude about the relevance of the proposed theoretical model for the description of carriers photo-generation and recombination in 2D-TMD monolayers. Obtained results demonstrate the possibilities of 2D-TMD quantum yield control by the gate voltage and indicate that 2D-TMDs are promising candidates for modern optoelectronics devices.

**Author's contribution.** M.V.S. formulated the problem, performed all analytical calculations and wrote the manuscript draft. A.N.M. and A.I.K. performed numerical calculations. A.N.M. compared with experiment, generated figures and improved the text.

**Acknowledgments.** A.N.M. is grateful to Eugene A. Eliseev (NASU) for the assistance in processing and fitting of experimental results.

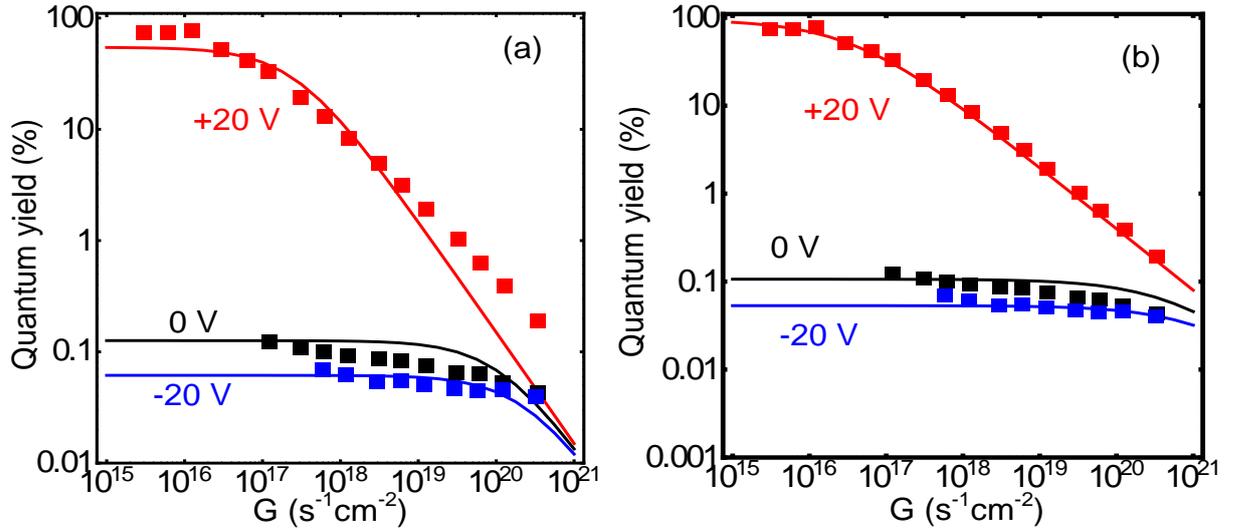

**Fig. A1.** (a) Symbols are experimental data from [17] for the gate voltages +20 V (red), 0 (black) and -20 V (blue), solid curves are calculated from Eq.(12) for fitting parameters **(a)** $V_T$= 19 V, $C_{ox}$=20 nF/m$^2$, $n_i = 10^8$ m$^{-2}$, $A$=15 SI units, $\alpha$=6.7×10$^8$ m$^2$ and $\beta$=50. **(b)** Solid curves are calculated from Eq.(15) for fitting parameters $V_T \approx$ 20 V, $C_{ox}$=15 nF/m$^2$, $N_{ex}$=1×10$^9$ m$^{-2}$, $m$=0.7, $N_T \left( \dfrac{\tau_{eff}}{N_0} \right)^m =$ 2.5×10$^{-3}$ SI units.